\documentclass[journal]{IEEEtran}
\usepackage[utf8]{inputenc}
\usepackage{amssymb}

\usepackage{cite}

\usepackage[pdftex]{graphicx}

\usepackage{amsmath}
\begin{document}

\title{FPGA code for the data acquisition and real-time processing prototype of the ITER Radial Neutron Camera}

\author{Ana~Fernandes, Nuno~Cruz, Bruno~Santos, Paulo~F.~Carvalho, Jorge~Sousa, Bruno~Gonçalves, Marco~Riva, Fabio~Pollastrone, Cristina~Centioli, Daniele~Marocco, Basilio~Esposito, Carlos~M.B.A.~Correia and Rita~C.~Pereira%
\thanks{Manuscript received June day, 2018; revised month day, 2018. 
The work leading to this publication has been funded partially by Fusion for Energy under the Contract F4E-FPA-327. IST activities also received financial support from “Fundação para a Ciência e Tecnologia” through project UID/FIS/50010/2013. This publication reflects the views only of the author, and Fusion for Energy cannot be held responsible for any use which may be made of the information contained therein.}
\thanks{A. Fernandes, N. Cruz, B. Santos, P.F. Carvalho, J. Sousa, B. Gonçalves, R.C. Pereira, are with Instituto de Plasmas e Fusão Nuclear, Instituto Superior Técnico, Universidade de Lisboa, 1049-001 Lisboa, Portugal (e-mail:anaf@ipfn.tecnico.ulisboa.pt).}
\thanks{M. Riva, F. Pollastrone, C. Centioli, D. Marocco, B. Esposito are with ENEA C. R. Frascati, Dipartimento FSN, via E. Fermi 45, 00044 Frascati (Roma), Italy.}
\thanks{C.M.B.A. Correia is with LIBPhys-UC, Department of Physics, University of Coimbra, P-3004 516 Coimbra, Portugal.}
}

\maketitle

\begin{abstract}
The main role of the ITER Radial Neutron Camera (RNC) diagnostic is to measure in real-time  the plasma neutron emissivity profile at high peak count rates for a time duration up to 500 s. Due to the unprecedented high performance conditions and after the identification of critical problems, a set of activities have been selected, focused on the development of high priority prototypes, capable to deliver answers to those problems before the final RNC design. This paper presents one of the selected activities: the design, development and testing of a dedicated FPGA code for the RNC Data Acquisition prototype. The FPGA code aims to acquire, process and store in real-time the neutron and gamma pulses from the detectors located in collimated lines of sight viewing a poloidal plasma section from the ITER Equatorial Port Plug 1. The hardware platform used was an evaluation board from Xilinx (KC705) carrying an IPFN FPGA Mezzanine Card (FMC-AD2-1600) with 2 digitizer channels of 12-bit resolution sampling up to 1.6 GSamples/s. The code performs the proper input signal conditioning using a down-sampled configuration to 400 MSamples/s, apply dedicated algorithms for pulse detection, filtering and pileup detection, and includes two distinct data paths operating simultaneously: i) the event-based data-path for pulse storage; and ii) the real-time processing, with dedicated algorithms for pulse shape discrimination and pulse height spectra. For continuous data throughput both data-paths are streamed to the host through two distinct PCIe x8 Direct Memory Access (DMA) channels. 
\end{abstract}


\IEEEpeerreviewmaketitle

\section{Introduction}
The ITER Radial Neutron Camera (RNC) main goal is to measure the plasma neutron emission profile enabling real-time plasma control purposes \cite{marocco}. Spectrometers, expected to be placed at the end of each collimated Line-Of-Sight (LOS), will provide the line-integrated neutron flux measurements for neutron emissivity calculations through inversion algorithms \cite{nuno}. A set of high-priority activities within the framework contract focus on the development of experimental setups of the neutron detector prototypes and its signal read-out equipment. This includes the design, development and testing of dedicated FPGA codes for the front-end electronics prototype \cite{marco}, aiming to acquire, process and store in real-time the incoming neutron and gamma fluxes at an expected sustained event rate of 2 MHz \cite{rita1}. The hardware platform includes an evaluation board from Xilinx (KC705) carrying an IPFN FPGA Mezzanine Card (FMC-AD2-1600) with 2 digitizer channels of 12-bit resolution sampling up to 1.6 GSamples/s \cite{rita2}. This paper presents the FPGA codes and algorithms of the front-end electronics prototype followed by some results achieved so far. 

\section{FPGA code}

\begin{figure}[!h]
	\centering
	\includegraphics[width=0.5\textwidth]{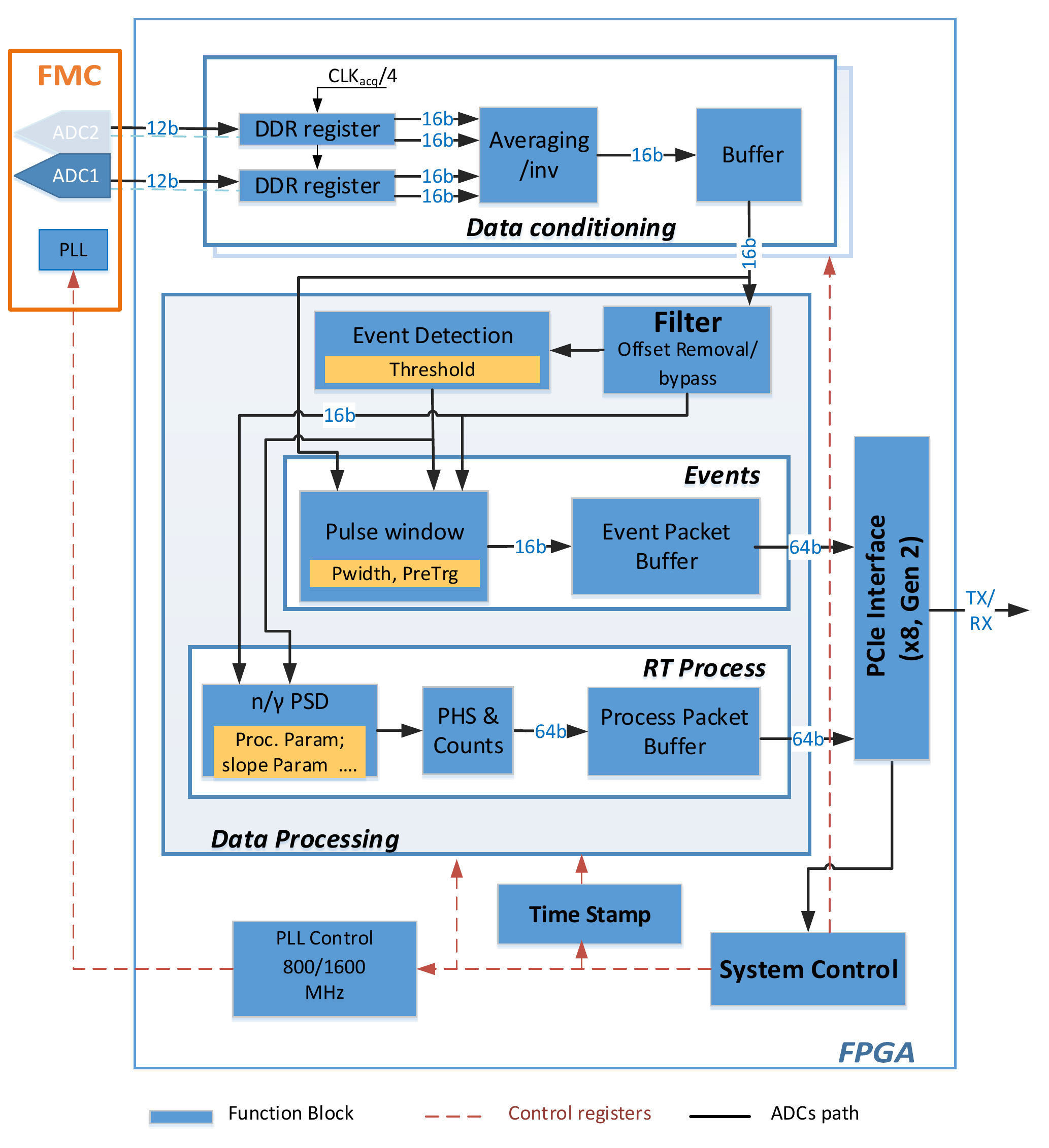}
	\caption{FPGA code flowchart of the RNC FPGA common environment and processing codes.}
	\label{fig:FPGACode}
\end{figure}

The fig. \ref{fig:FPGACode} flowchart depicts the main blocks of code developed under the RNC framework contract concerning the FPGA common environment and algorithm activities. The RNC project was develop in Verilog with the Xilinx VIVADO tool (2015.4 and 2017.4 versions), implemented in the Xilinx KC705 + IPFN FMC prototype, and tested using synthetic pulses from CAEN generator (DT5800D). According with fig. \ref{fig:FPGACode}, the code is composed of four main blocks, detailed in the next sections: i) data conditioning; ii) data processing; iii) data streaming (PCIe interface); and iv) system control.  

\section{Data conditioning}

The RNC detector signals are digitized by the 12-bit, 1.6 Gsamples/s Analog to Digital Converters (ADC) of the FMC card. The ADCs are configured (e.g. sampling rate, operating mode) by FPGA (PLL control module, (fig. \ref{fig:FPGACode}), through the 11 24-bit registers of the high performance frequency synthesizer (LMX2531) with Phase Locked Loop (PLL) installed in the FMC module.   
The ADCs are programed to operate in the called demux mode, where data from the analogue input are produced at half of the acquired rate, at twice of the number of output buses. Thus, the FPGA receives two differential Double Data Rate (DDR) 12-bit buses per ADC, with data sampled at 1/4 of the acquisition rate (400 Msamples/s). The signal conditioning block includes Input DDR (IDDR) primitives, used to retrieve data from both clock edges, resulting in a total of 4 samples per sampling clock (1/4 of the acquisition clock). This block outputs the average of each four samples acquired in the same sampling clock, increasing the ADC Effective Number Of Bits (ENOB) resolution in one bit. 

\section{Data processing}

Considering the ultra-high acquisition rates allowed by the FMC card (1.6 Gsamples/s), dedicated data reduction and processing algorithms are a priority for feasible data streaming. Dedicated algorithms should be able to sustain the expected high data throughput (0.5 GB/s per ADC) without losses. Thus, two different operating modes were selected: i) Events: the detected events are streamed to host together with the corresponding time occurrence, Time Stamp (TS), for data archiving; ii) Real-Time process: delivers the gamma-ray/neutron Pulse Shape Discrimination (PSD) and/or Pulse Height Spectra (PHS) in real-time. The next subsections describe the main real-time algorithms of the data processing block.

\subsection{Filter}
\label{Filter}

When signals coming from detectors are unstable (e.g. drift in the signal baseline), the pulse detection may fail, leading to a poor system performance (e.g. energy resolution degradation of founded events) \cite{caen}. Dedicated pulse filtering algorithms may be applied to raw data before the event detection stage, capable of digitally restore the baseline and remove undesired offset. The optimal offset removal/baseline restoring algorithm usually depends on the incoming signal on-site after diagnostic installation. Thus it is not possible to select at this phase the best algorithm capable to minimize possible instabilities in the RNC signals. However, a generic filter module interface was added to the project (FILTER, fig. \ref{fig:FPGACode}) enabling the possibility to allocate suitable stabilization circuits. The filter module can be bypassed if not needed. The event detector algorithm may trigger from filtered data, when filter is ON, or from raw data when the filter bypass option is selected. 
As example, a Digital Trapezoidal based Shaper (DTS) was implemented in the filter module. DTS is a well known technique capable of suppressing ballistic deficit of sharp peak with exponential decay pulses, being a strong candidate for baseline restoring and offset removal of RNC signals \cite{caen},\cite{trapezoidal}. When the DTS based filter is used, the exponential signal is transformed into a kind of trapezoid, whose amplitude is proportional to the energy of the event \cite{ana_algorithms}. Considering expected pileup, DTS parameters were slightly modified providing filtered events similar to Gaussians instead of pure trapezoids \cite{ana_2018}.

\subsection{Event detection}

Considering the expected shape of gamma-ray/neutron from RNC scintillators (exponential decay signals with fast rise time), two different trigger types were selected for event detection: i) Basic Trigger: threshold by level. The event detector triggers when data reaches a predefined threshold; ii)	Advanced Trigger: threshold by derivative. The event detector triggers when first signal derivative reaches a predefined threshold. The advanced trigger is the option adopted by many spectroscopy diagnostics due to its ability to reject the high frequency noise, baseline restoring, and cancel the low frequency fluctuations \cite{caen}. The event detection algorithm is applied to both raw data and filtered data.  

\subsection{Events storage - Pulse Window}

\begin{figure}[!h]
	\centering
	\includegraphics[width=0.4\textwidth]{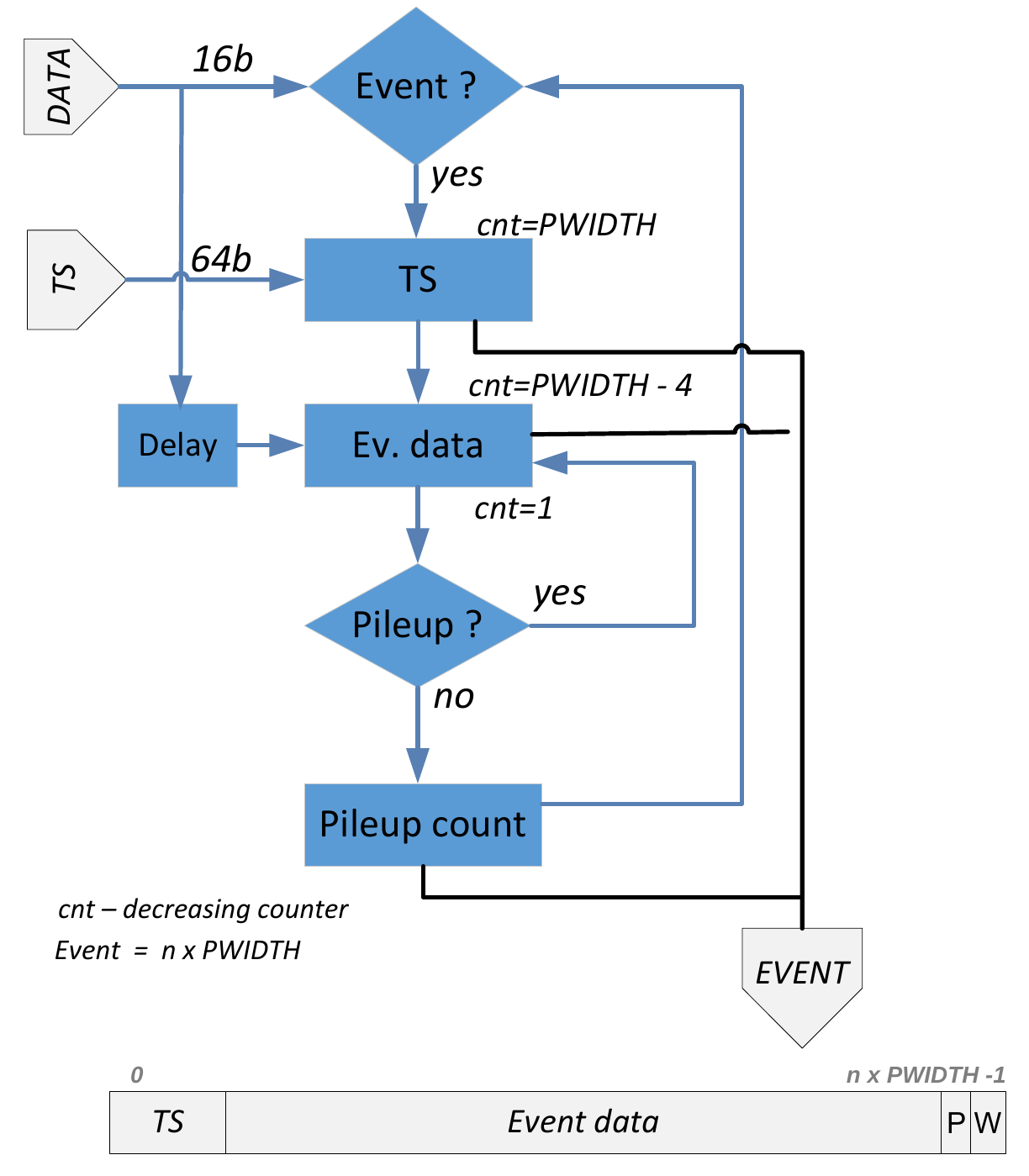}
	\caption{Event detection and storage flowchart. The number of PWIDTH stored depends on pileup occurrences in the second half of each PWIDTH.}
	\label{fig:event}
\end{figure}

The event based module receives data from ADCs, aligned with the trigger event detector. When an event is detected, the FPGA starts storing samples (16-bit) in a pre-defined Pulse Window (PWIDTH), including the Pre-Trigger samples (PTRG) required to provide the baseline level. The number of samples corresponding to each stored event depends on pileup occurrences. If a new event is detected during the second half of the PWIDTH being stored, a new PWIDTH is stored, as described by fig. \ref{fig:event} flowchart. Thus, each event is composed by the TS value (64-bit), followed by n x PWIDTH samples (excluding the last 32-bit, explained later), corresponding to the event data. 
The module state machine includes increasing counters to provide the number of triggers occurred in each event (pileup detection) and the number of PWIDTH used. Thus, the event window ends with the counter value corresponding to the total number of pulses (P) found per event (16-bit), followed by the number of extra PWIDTH (PWIDTH-1) used to store the event (8-bit), and finally by an end-of-event (W) tag (8-bit). The event data follow to an event packet buffer, a two domains clock First-In-First-Out (FIFO) buffer, for data streaming.

\subsection{Real-time process - PSD}
\label{Real-time process - PSD}

\begin{figure}[!h]
	\centering
	\includegraphics[width=0.4\textwidth]{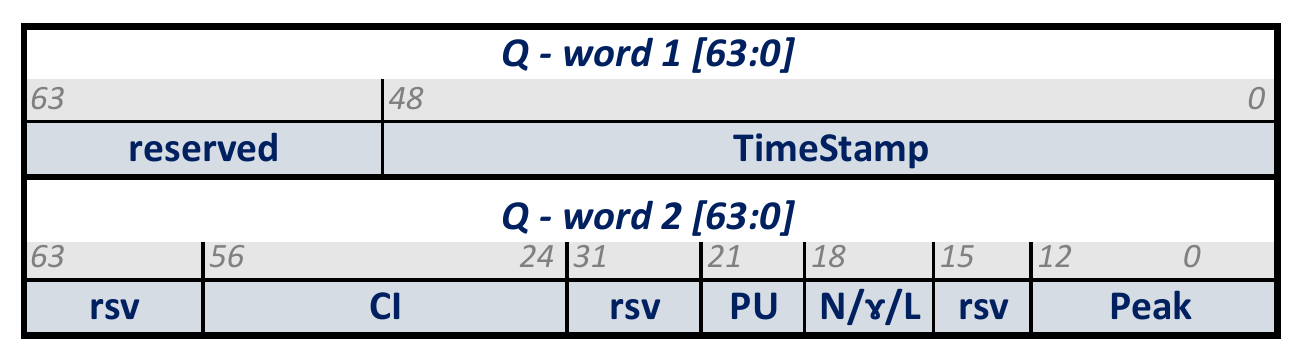}
	\caption{PSD output data packet. Two Q-Word (64-bit) with peak and CI values, particle type (neutron/gamma/LED), pileup (PU) and corresponding TS of each detected event)}
	\label{PSD_packet}
\end{figure}

Different algorithms, feasible to implement in FPGA, can be used to perform neutron/gamma discrimination \cite{marco_JET}, \cite{bipolar}. Similar to the filter module (sec. \ref{Filter}), a generic PSD interface was implemented, foreseeing user defined inputs capable to meet different algorithm needs (e.g. calibration slope and event type parameters). For testing, it was implemented a PSD code based on DTS, receiving as input data from filter module. The neutron/gamma discrimination is determined by the relation factor between the maximum of the trapezoid (peak value), and the trapezoid area - charge integration (CI) \cite{rita_psd}. From this relation it is possible to determine if the detected event is neutron or gamma, depending if the result is below or above of the corresponding calibration slope value. The foreseen Light Emission Diode (LED) detection, for calibration purposes, was not included in this implementation. However, due to its singular shape, the LED detection is not a critical concern. The data packet returned by the PSD module, fig. \ref{PSD_packet}, is a two Q-Word (2 x 64-bit) containing the PSD output of each detected event. 

The PSD output data may be used to feed the PHS module (sec. \ref{Real-time process - PHS}), when available, or streamed directly to host.  When the PHS module is present, the slope parameters must be previously adjusted for proper particle discrimination and correct PHS construction at FPGA.
 
\subsection{Real-time process - PHS}
\label{Real-time process - PHS}

\begin{figure}[!h]
	\centering
	\includegraphics[width=0.5\textwidth]{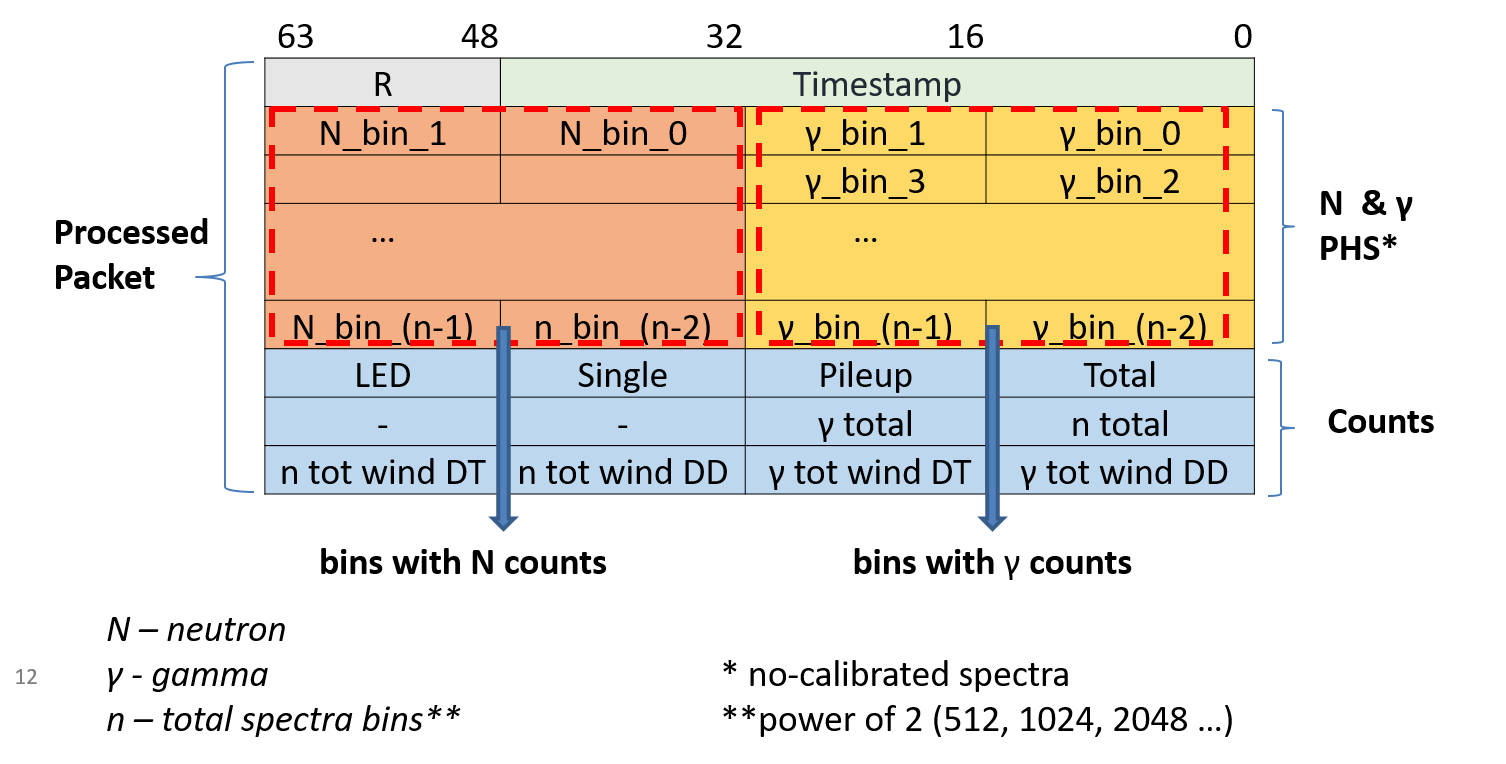}
	\caption{Real-time packet with both neutron and gamma PHS, and count values, to be streamed periodically to host.}
	\label{fig:phs_packet}
\end{figure}

The PHS module receives the two PSD Qwords, being responsible for the real-time PHS construction of both neutron and gammas. For each real-time cycle, established by the Synchronous Data Network (SDN) periodicity \cite{rita1}, a data packet (fig. \ref{fig:phs_packet}) with both uncalibrated spectra and the corresponding counts (number of single and piled-up events; neutron, gamma, LED and total counts; counts per bin window) is streamed to host. The state machine of the PHS and counts module is responsible for: i) de-serialization of the two PSD Qwords; ii) selection of the pre-defined value for spectra construction (peak or CI); ii) founding the corresponding histogram address for each incoming neutron/gamma; iii) counters incrementation; iv) packet construction and storage in buffer. In each SDN cycle a new PHS packet is streamed to host and buffer reset (e.g. 2ms). 

\section{Data streaming}

The 8-lane PCIe (5 GT/s, Gen2) was the selected communication protocol for data transfer between RNC prototype and its host. Both RNC events and real-time processed data are streamed to host through dedicated PCIe Direct Memory Access (DMA) packets, resulting in higher data throughput and better overall system performance through lower CPU utilization \cite{rita2}. 
Two distinct DMA channels were implemented for data streaming: i) DMA 0 to carry the events for data archiving/host processing; ii) DMA 1 for the real-time processed data (PSD or PHS packets). 

Considering the RNC demanding data transfer at variable rate (maximum throughput of 4 Mevents/s per FMC) a third DMA (DMA 2) was included to stream the status information (e.g. last sent DMA, DMA 0/1 counters, last DMA address), detected by host through a pooling mechanism. Thus, each streamed DMA 0/1 data-set written to host, is followed by the DMA 2 carrying a new status word. This allows to identify the last DMA data transfer for proper data retrieval from host memories. Usually, in less demanding applications, the status word is written in PCIe shared memory as completion to a CPU read request (sec. \ref{System control}). This procedure may enable conflicts in the PCIe bus between the register reading and DMA data transfer, which are of higher probability for demanding throughput at variable rate.

\begin{figure}[!h]
	\centering
	\includegraphics[width=0.5\textwidth]{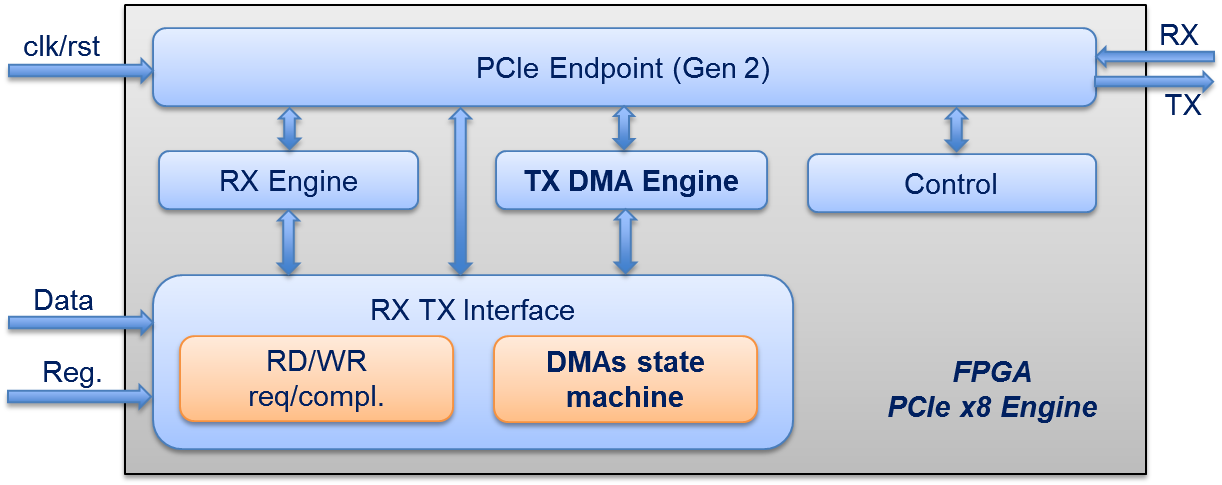}
	\caption{PCIe x8 (GEN 2) interface flowchart.}
	\label{fig:PCIe}
\end{figure}

The DMA engine is included in PCIe Receiving (RX)/ Transmition (TX) - interface (RX-TX), where a state machine is responsible for data-path management between RX and TX engines (endpoint front-end interfaces) and other FPGA modules, as depicted in fig. \ref{fig:PCIe} flowchart.

\section{System control}
\label{System control}

The endpoint configuration is done through shared configuration registers, located in the host shared memory namely PCIe Base Address (BAR), usually settled by host BIOS during PCIe configuration space at power up. Registers must be properly defined by host (driver) and endpoint (FPGA code) guaranteeing its correct operation. The System Control module interfaces with the PCIe engine, receiving and delivering 32-bit register fields to/from host. Moreover it exchanges configuration registers with other FPGA modules.

\section{Results}

\begin{figure}[!h]
	\centering
	\includegraphics[width=0.4\textwidth]{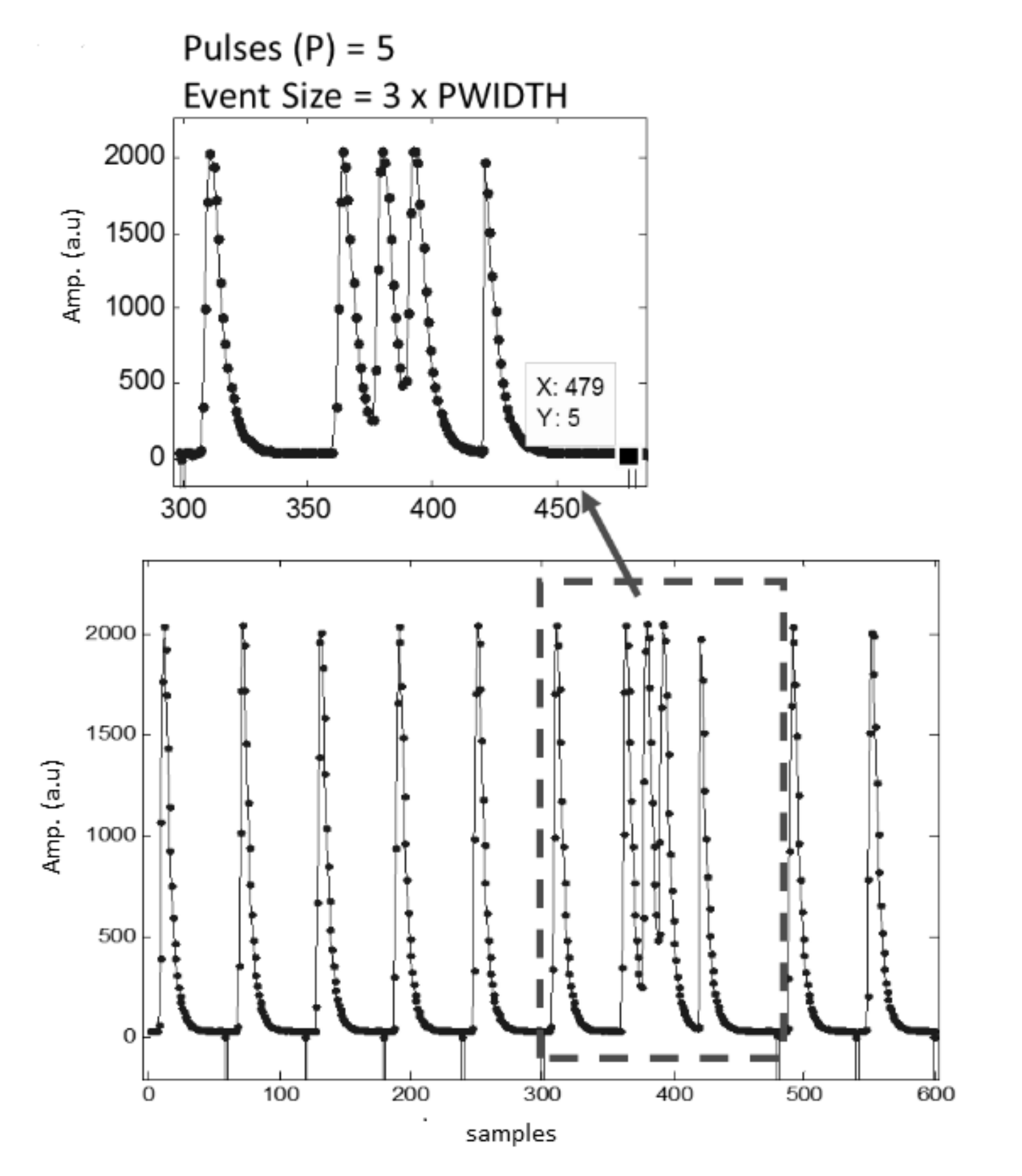}
	\caption{Sequence of events stored with RNC prototype using synthetic data from CAEN emulator. Zoomed picture: event window composed of 3 PWIDTH (3 x 64-samples) and 5 piled-up pulses, identified by the corresponding P value (y=5).}
	\label{fig:event_results}
\end{figure}

The RNC FPGA code was tested using synthetic data from CAEN DT5800D pulse emulator. As example, Fig. \ref{fig:event_results} shows a sequence of events from DMA 0 data streaming. Zoomed fig. highlights the sliding event window and pileup detection capabilities.

\begin{figure}[!h]
	\centering
	\includegraphics[width=0.4\textwidth]{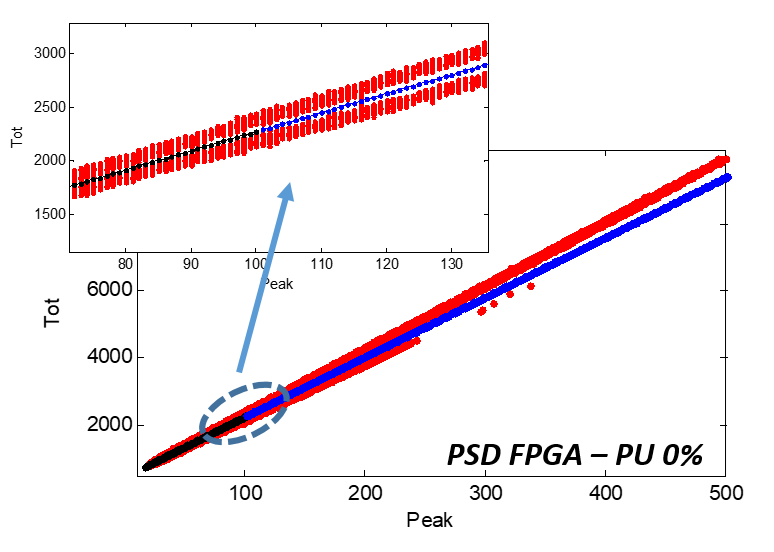}
	\caption{Relation factors (red) used for neutron/gamma discrimination and the corresponding separation slopes (black and blue) included by an offline code.}
	\label{fig:psd}
\end{figure}

In fig. \ref{fig:psd} it is possible to observe the well separated relation factors in red (x: peak; y: Tot (CI)) from PSD packet directly streamed to host through DMA 1. To simulate neutron and gamma events it was used the CAEN emulator providing synthetic gamma and neutron shaped pulses from two combined channels. RNC prototype receives data at an event rate up to 1 Mevents/s from both channels simultaneously, without pileup. The separation slopes (blue and black) are included for better data analysis using an offline matlab code.   

\begin{figure}[!h]
	\centering
	\includegraphics[width=0.5\textwidth]{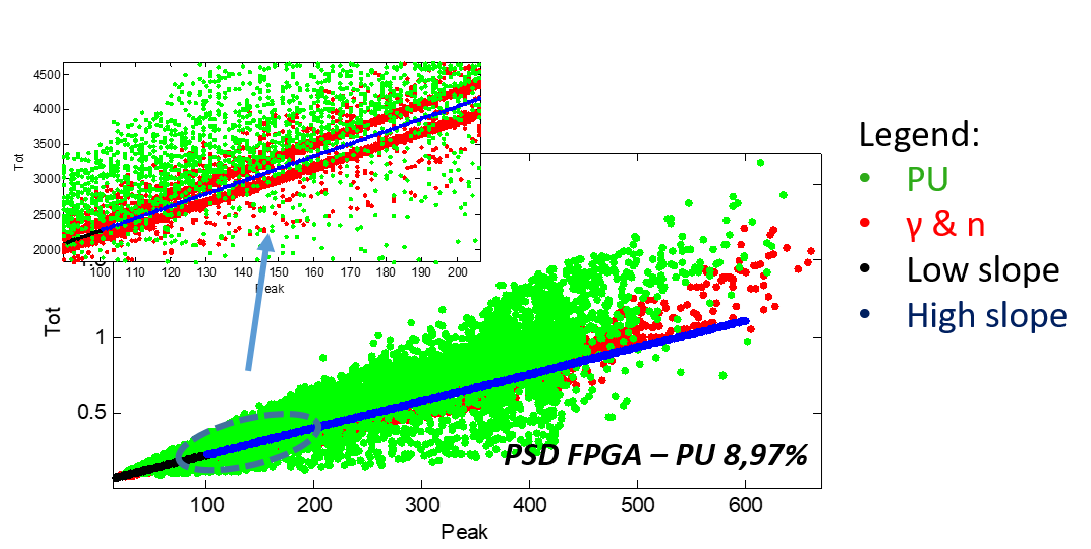}
	\caption{FPGA PSD outputs with the pileup detection highlighted (green), when the Poisson distribution was applied to n based-shape channel.}
	\label{fig:psd-pu}
\end{figure}

To test if the FPGA processing code is capable to identify piled-up events it was applied a Poisson distribution individually (10 \% of pileup) to each CAEN channel. As example fig. \ref{fig:psd-pu} depicts the piled-up events found by PSD at FPGA (green spots) superimposed with the well-defined neutron/gamma relation factor. 

From experimental results it was concluded that the FPGA algorithms are feasible to detect piled-up events in both event and processed data. However it was observed that pileup detection slightly reduces for filtered data. This is explained by the signal smoothing effect imposed by the DTS filter. Improvements in the event detection algorithm where identified for future implementation (e.g. maximum pulse length), capable to reduce the undesired smoothing effect. 

\begin{figure}[!h]
	\centering
	\includegraphics[width=0.4\textwidth]{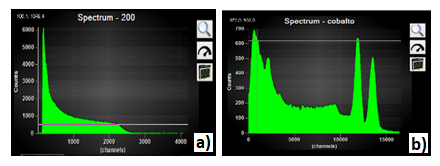}
	\caption{Input spectra of CAEN emulator simulating single neutron and gamma spectrum. a) gamma based-shape; b) neutron based-shape pulses.}
	\label{fig:caen_spectra}
\end{figure}

\begin{figure}[!h]
	\centering
	\includegraphics[width=0.5\textwidth]{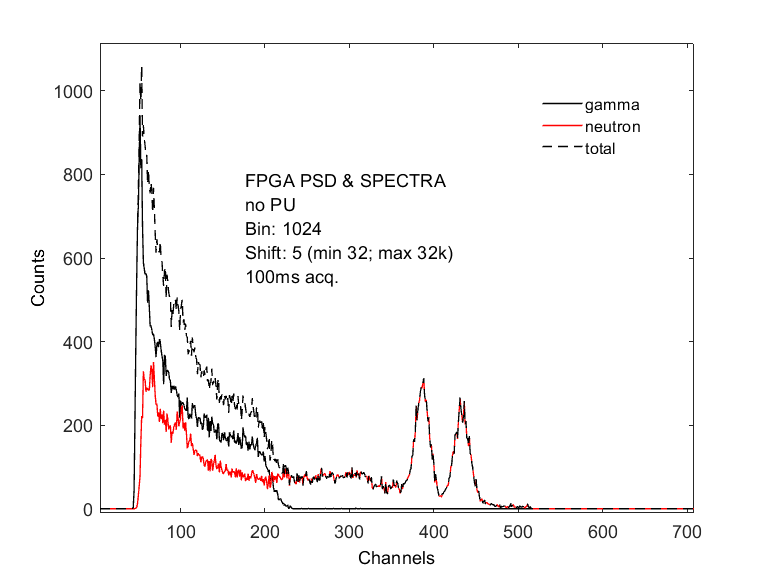}
	\caption{Real-time PHS from FPGA (neutron, gamma and total spectra), in agreement with the expected results.}
	\label{fig:phs}
\end{figure}

To check the FPGA PHS algorithm performance, streamed through DMA 1, the real-time PHS were compared with spectra obtained by post processing the event data from DMA 0 acquired simultaneously. As example fig. \ref{fig:phs} depicts the resulting PHS from FPGA  for a 100 ms acquisition using two CAEN emulator combined channels (Ch1: 500kev/s of gamma based-shape pulses through an input spectrum defined by fig. \ref{fig:caen_spectra} a); Ch2: 500 kev./s of neutron based-shape pulses through an input spectrum defined by fig. \ref{fig:caen_spectra}b)). Please note that both spectra physics is meaningless.   

It was concluded that the FPGA PHS output, depicted in fig \ref{fig:phs} is in agreement with input data (fig. \ref{fig:caen_spectra}), and with spectra from post processing methods, using event data from DMA 0. 

However, before operating the FPGA PHS module, it is necessary to deeply adjust the separation slope values needed for proper discrimination. Thus the real-time PHS production at FPGA might be difficult to use in experiments with higher fluctuation of the separation slopes. In these cases the best option is to stream the PSD data (peak and CI relation factors) through DMA 1, and build the PHS at host.

\section{Conclusion}

This paper presents the FPGA code developed for RNC front end electronics prototype. The code foresees to acquire, process and store in real-time the neutron and gamma events from the detectors located in collimated LOS viewing a poloidal plasma section. The code was implemented and tested in an evaluation board from Xilinx (KC705) carrying an IPFN FPGA Mezzanine Card (FMC-AD2-1600) with 2 digitizer channels of 12-bit resolution sampling up to 1.6 GSamples/s. After signal condition, the code uses dedicated algorithms for event detection, filtering, pileup detection and real time processing (event storage, PSD and PHS). Three distinct x8 GEN2 DMA channels were implemented. Two DMAs are responsible for real-time data streaming (event-based and PSD or PHS processed data), and the third DMA for the status word, avoiding the concurrent access of reading requests. The code was successfully tested with synthetic data from a CAEN emulator in laboratory, allowing a maximum throughput of 1600 MB/s (the maximum possible for 2 channels @ 400 MHz – continuous acquisition). Concerning the PSD algorithm, it is possible to conclude that PSD relation factors from FPGA provide successful neutron/gamma discrimination, when compared with post processing methods. However it was observed that pileup detection slightly reduces for filtered data, when compared with post processed event data from the same acquisition. This is due to the signal smoothing effect introduced by the DTS filter. New methods were identified capable to overcome this undesired effect in presence of pileup. The PHS algorithm was also successfully implemented and tested at FPGA, which results are in agreement with post processing PHS using event data. However it was concluded that PHS at FPGA might be unfeasible in experiments with higher fluctuation of the separation slopes (e.g. signal gain changes). Thus, the feasible solution is to stream the PSD packets through DMA 1 for PHS construction at host.

\section*{Acknowledgment}

This manuscript is in memory of Professor Carlos Correia who is no longer among us.

\end{document}